\documentclass[final]{aipproc}

\layoutstyle{6x9}

\newcommand{\etal}{et\,al.\ }
\newcommand{\logg}{\mbox{$\log g$}}
\newcommand{\Teff}{\mbox{$T_\mathrm{eff}$}}

\newcommand{\lppr}{\stackrel{<}{\scriptstyle \sim}}
\newcommand{\lappr}{\raisebox{-0.4ex}{$\lppr $}}

\newcommand{\ion}[2]{\mbox{#1\,{\sc #2}}}
\def\kpd{KPD\,0005$+$5106}
\def\hh{H1504$+$65}

\begin{document}

\title{HST/COS Spectroscopy of H1504+65}

\classification{97.20.Rp, 97.10Ex, 97.10Tk}
\keywords      {White dwarfs, atmospheres, abundances}

\author{Klaus Werner}{
  address={Institute for Astronomy and Astrophysics, Kepler Center for
  Astro and Particle Physics, University
  of T\"ubingen, 72076 T\"ubingen, Germany}
}

\author{Thomas Rauch}{
  address={Institute for Astronomy and Astrophysics, Kepler Center for
  Astro and Particle Physics, University
  of T\"ubingen, 72076 T\"ubingen, Germany}
}

\begin{abstract}
We present new ultraviolet spectra of the peculiar white dwarf (WD) \hh,
obtained with \emph{COS} on \emph{HST}. \hh\ is the hottest known WD (\Teff\ =
200\,000\,K) and has an atmosphere mainly composed of C and O, augmented with
high amounts of Ne and Mg. This object is unique and the origin of its surface
chemistry is completely unclear. We probably see the naked core of either a C--O
WD or even a O--Ne--Mg WD. In the latter case, this would be the first direct
proof that such WDs can be the outcome of single-star evolution. The new
observations were performed to shed light on the origin of this mysterious
object.
\end{abstract}

\maketitle

\section{Introduction}

\hh\ is a faint blue star that was identified as the counterpart of a bright
soft X-ray source \citep{nousek:86}. Spectroscopically, it is a member of the
PG\,1159 class but, within this class \hh\ is an extraordinary object. It was
shown that it is not only hydrogen-deficient but also helium-deficient.  From
optical spectra it was concluded that the atmosphere is primarily composed of
carbon and oxygen, by equal amounts \citep{werner:91}. Strong neon lines were
detected in soft X-ray spectra taken with the \emph{EUVE} satellite and in a
Keck spectrum, and a high abundance of neon was derived \citep{werner:99}. The
origin of this exotic surface chemistry (C = 49\%, O = 49\%, Ne = 2\%, mass
fractions) is completely unclear. We have speculated that \hh\ represents the
naked C--O core of a white dwarf. Another, even more exciting possibility is
that we see the eroded C--O envelope of a O--Ne--Mg white dwarf. This is
corroborated by our analysis of a \emph{Chandra} soft X-ray spectrum
\citep{werner:04}, which allowed the detection of magnesium.

\hh\ turns out to be the hottest single WD ever analyzed with model atmospheres
(\Teff\ = 200\,000\,K $\pm$ 20\,000\,K, \logg\ = 8.0 $\pm$ 0.5), only rivaled by
the hot DO \kpd\ (see Wassermann \etal in these proceedings, and
\cite{was10}). Because of this extremely high temperature, a unique photospheric
absorption-line spectrum can be observed with \emph{Chandra-LETG}. It shows a
wealth of lines from highly ionized species (\ion{O}{vi}, \ion{Ne}{vi} --
\ion{Ne}{viii}, \ion{Mg}{v} -- \ion{Mg}{viii}). But the spectral analysis of the
X-ray data is seriously hampered by heavy line blanketing of iron-group
elements. In principle, we can account for this in our synthetic spectra, but two
problems prevent a detailed  quantitative analysis of relatively weak absorption
lines from light metals within the Fe-group forest. First, accurate line
positions are unknown for the majority of the Fe-group lines. Second, the bulk
of the Fe-group lines from very high ionization stages (a single \ion{Fe}{x} UV
line was recently discovered in a \emph{FUSE} spectrum, see Werner, Rauch, \&
Kruk, these proceedings; and \cite{we10}) are completely unknown. Consequently,
effects of Fe-group line blending on other weak metal lines cannot be calculated
with sufficient precision. Nevertheless, we have roughly estimated the Mg
abundance to about 1\%, an amount similar to the Ne abundance. If this strong
overabundance (20 times solar) can be confirmed, then that would strongly
support the idea that \hh\ is a O--Ne--Mg WD.

This means that  \hh\ might had been one of the ``heavy-weight''
intermediate-mass stars (8\,M$_{\odot}\ \lappr$ $ M\ \lappr$\ 10\,M$_{\odot}$)
that form white dwarfs with electron-degener\-ate O--Ne--Mg cores. Evolutionary
models \cite{iben:97} predict strong Ne and Mg overabundances in the C/O
envelope. Another strong argument in favor of this idea would be the detection
of sodium, which would be direct evidence for C-burning. The models predict that
the $^{23}$Na abundance at the bottom of the C/O envelope is comparable to that
of neon (main isotope $^{20}$Ne) and magnesium ($^{24,25,26}$Mg, see Fig.~34 in
\cite{iben:97}). Unfortunately, we were not able to detect Na lines in the
\emph{Chandra} spectrum beyond doubt because of, again, heavy metal-line
blanketing.

\section{Relevance of H1504+65}

At present it is uncertain under which circumstances super-AGB stars (i.e. the
massive counterparts of AGB stars that ignite carbon but do not proceed to
further stages of nuclear burning) produce O--Ne--Mg WDs or explode as
electron-capture SNe producing NSs (e.g. \cite{si07}). This uncertainty mainly
arises from modeling uncertainties in mass-loss and mixing processes.  The
possibility that \hh\ is a O--Ne--Mg WD is remarkable, because evidence for the
existence of such objects is rather scarce \citep{weidemann:03}. Evidence from
single massive WDs is weak, and the most convincing cases are WDs in binary
systems. Strong Ne overabundances are found in novae \citep{livio:94} or in
eroded WD cores in LMXBs \citep{juett:01}. If we can prove high Mg and Na
abundances in \hh, then this would be the most compelling case for the existence
of a single O--Ne--Mg WD, i.e. a post super-AGB star.

It is interesting to note that the recently discovered relatively cool white
dwarfs (\Teff\ $\approx$ 20\,000\,K, see Dufour et al.,  these proceedings, and
\citep{dufour:07}) with almost pure carbon atmospheres (so-called hot DQs), as
well as the O-rich white dwarfs (He-dominated atmospheres with
O/He\,$\approx$\,0.01 and O\,$>$\,C, by number, and \Teff\ around 10\,000\,K)
reported by G\"ansicke \etal (these proceedings, and \cite{gae10}), could be
evolutionary linked to \hh. The latter, in particular, could well be O--Ne--Mg
white dwarfs.

\hh\ also challenges stellar evolution theory relevant for super-AGB stars,
because it cannot explain how \hh\ has lost its H-rich and He-rich envelopes and
why it exposes its metal-rich core. Our detailed abundance determination of the
H-deficient PG\,1159 stars has shown that the efficiency of convective overshoot
during He-shell flashes is stronger than hitherto assumed in evolutionary
calculations. In analogy, the determination of the strange surface chemistry of
\hh\ challenges super-AGB evolutionary models, which are unable to explain the
observed helium-deficiency.

\begin{figure*}[th!]
\includegraphics[width=0.95\textwidth]{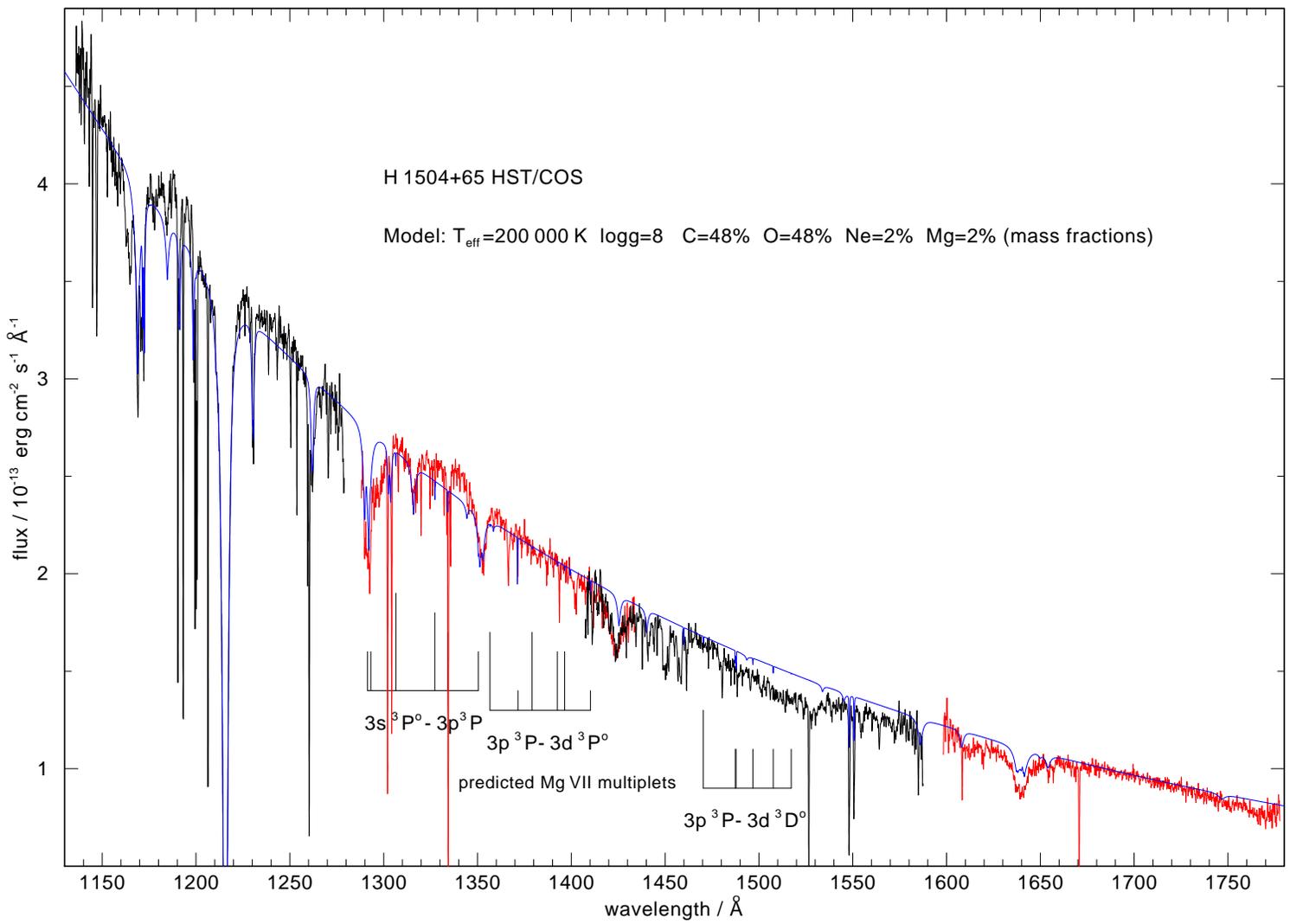}
\caption{\emph{HST/COS} observation of \hh\ compared to a
  model. Three \ion{Mg}{vii} multiplets, predicted by
  the model, are marked.}\label{fig1}
\end{figure*}

\begin{figure}[t]
\includegraphics[width=\columnwidth]{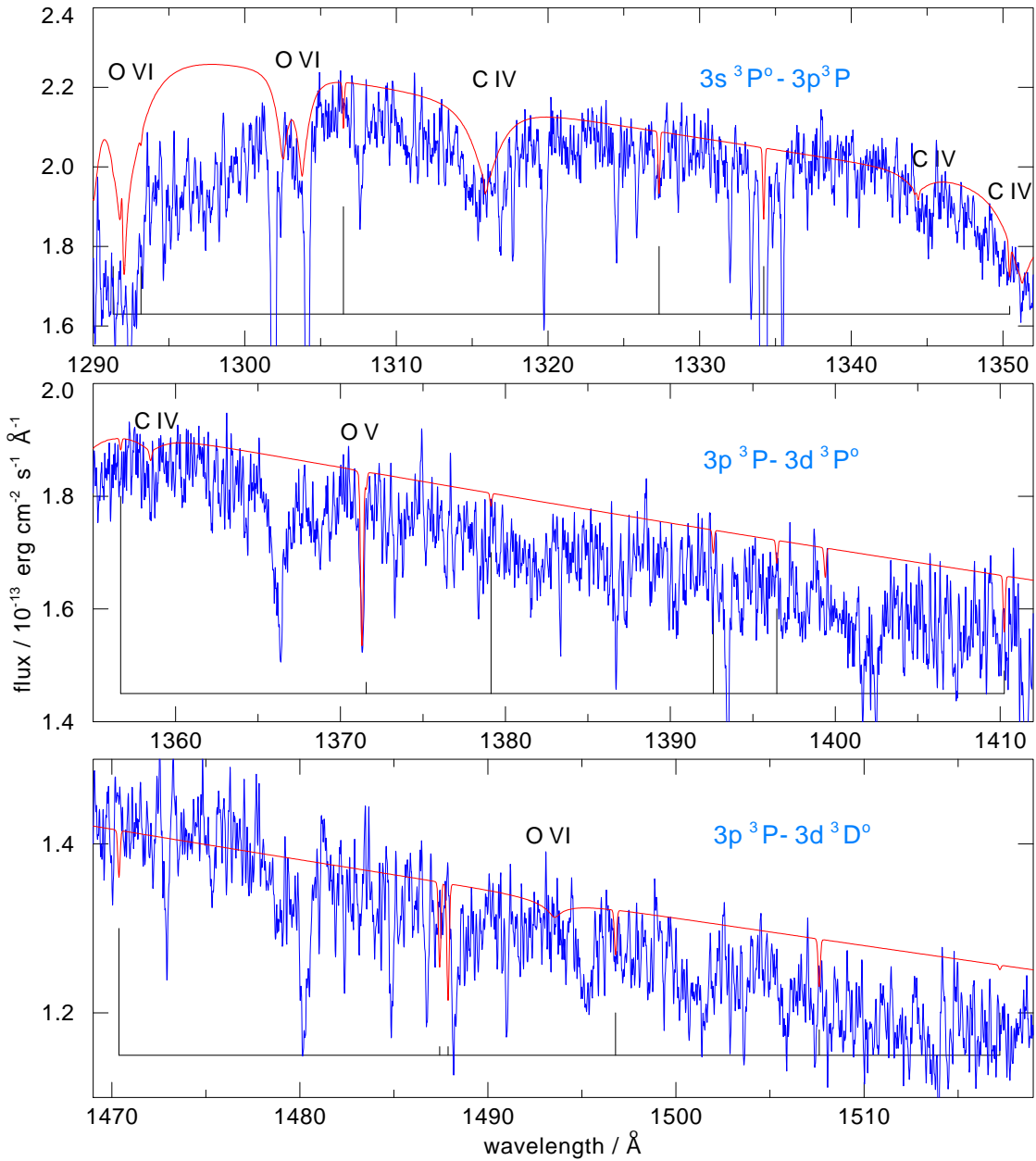}
\caption{Details of the spectra in the vicinity of the three
  \ion{Mg}{vii} multiplets marked in Figure~\ref{fig1}.}\label{fig2}
\end{figure}

\begin{figure}[t]
\includegraphics[width=\columnwidth]{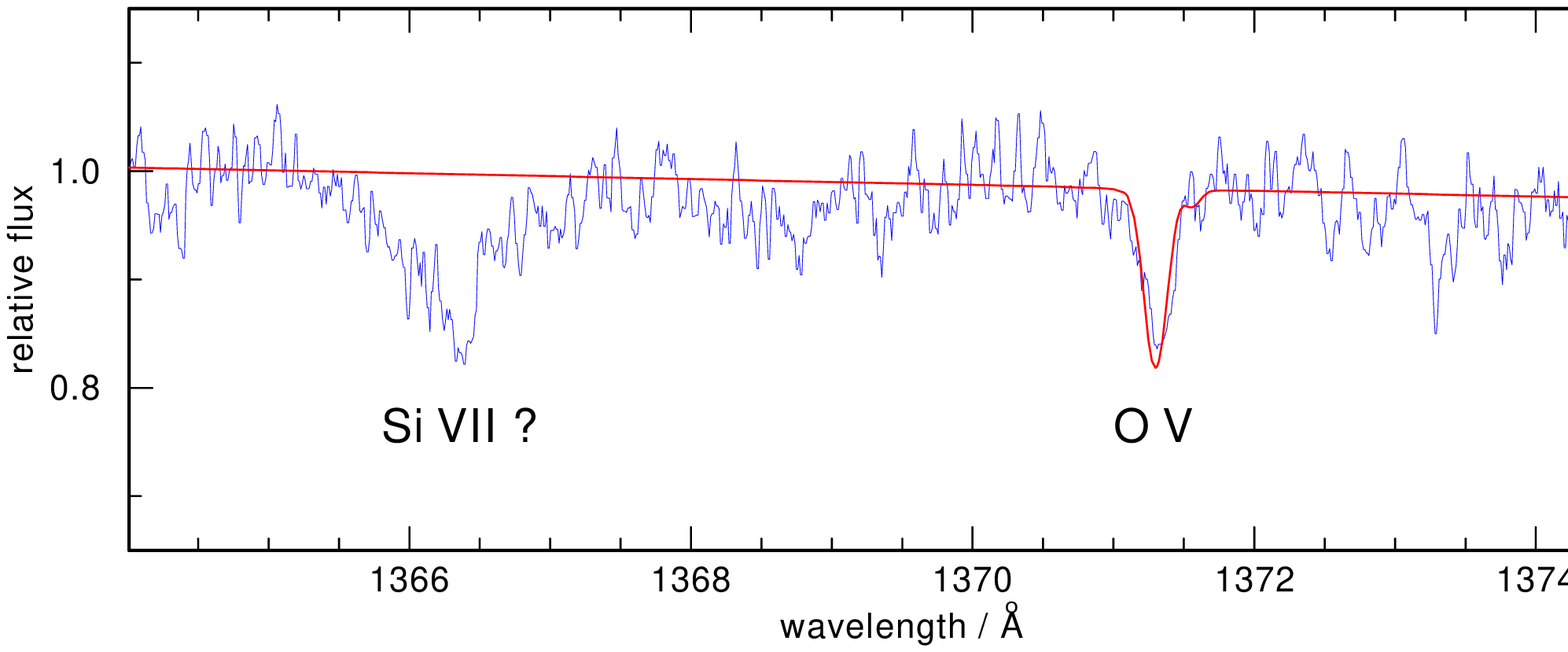}
\caption{Detail of the \emph{COS} and model spectra in the vicinity of the 
\ion{O}{v}~1371~\AA.}\label{fig3}
\end{figure}

\section{\emph{HST/COS} UV spectroscopy}

Our immediate motivation to perform new observations is the search for Mg and Na
lines in the UV spectrum of \hh\ and to determine the respective abundances. The
UV range is certainly free of iron-group lines. Hot central stars of planetary
nebulae (\Teff\ $\approx$ 100\,kK) show iron lines of \ion{Fe}{v} --
\ion{Fe}{vii} in the UV. But \hh\ is so hot that the dominant ionization stages
of the Fe-group elements are {\sc ix} and {\sc x}. Almost all respective lines are located
in the EUV and soft-X-ray regions. Hence, we should be able to detect and
analyze even very weak lines of light metals in the UV. We expect to see Mg and
Na lines if these elements are abundant on the 1\% level. There are three
multiplets of \ion{Mg}{vii} and two multiplets of \ion{Na}{vi} located in the
1150--1700~\AA\ region.

The \emph{Cosmic Origins Spectrograph (COS)} was installed at \emph{HST} in May
2009. On Oct.\ 24, 2009, \emph{HST} recovered from a shutdown of its science
data formatting (SIC\&DH) onboard computer, an event that spoiled our
observations because they were performed shortly afterwards, on the same day.
\emph{COS} spectroscopy of \hh\ was performed with gratings G130M and G160M
during one orbit each. The wavelength range 1150--1760~\AA\ was covered with
about 0.1~\AA\ resolution. Data inspection revealed problems with the flux
calibration, related to the recovery of SIC\&DH. It turned out that the
\emph{COS} detectors worked outside of their normal temperature range. As a
consequence, image distortions were large enough to place stim pulses --
required for the distortion correction -- off the detector. Our request for
repetition of the observations was accepted. This repetition was performed on
May 25, 2010.

\begin{figure}[t]
\includegraphics[width=0.84\columnwidth]{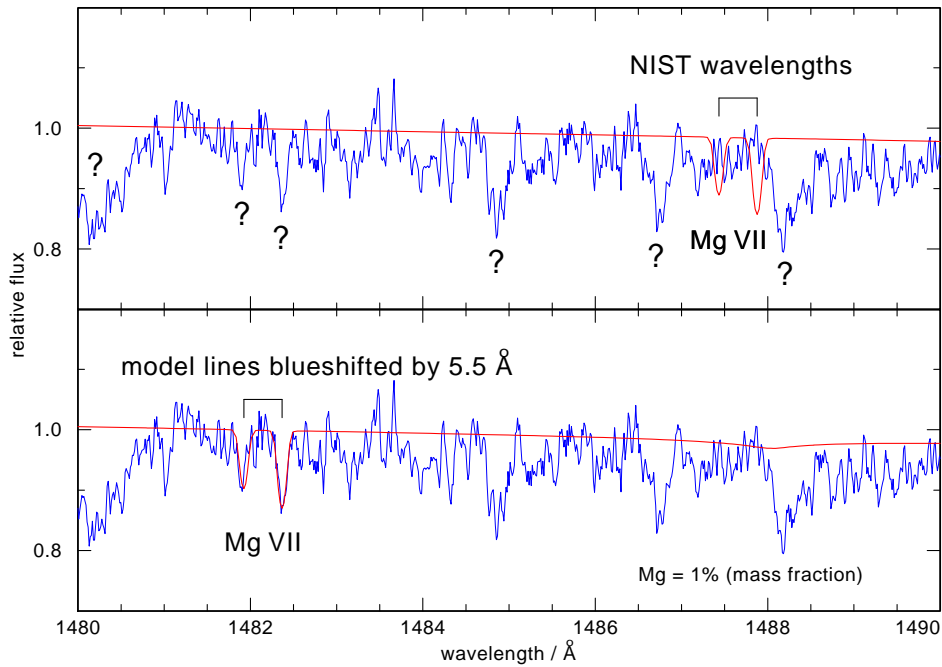}
\caption{Top: Detail of the \emph{COS}  spectrum in the vicinity of two \ion{Mg}{vii}
  lines marked in the bottom panel of Figure~\ref{fig2}. The predicted
  \ion{Mg}{vii} lines are not present in the observation. A number of
  unidentified photospheric lines are visible (question marks). Bottom: Computed spectrum
  shifted by 5.5~\AA.}\label{fig4}
\end{figure}

Figure~\ref{fig1} displays the four \emph{COS} spectral segments compared to a
spectrum computed from a model with parameters found from our earlier
multi-spectral analyses mentioned above. Almost all prominent line features are
from \ion{C}{iv}, \ion{O}{vi}, and \ion{Ne}{viii}, hence, elements (and ions)
that were already identified in \emph{FUSE} spectra \cite{werner:04}. New
identifications are lines from \ion{O}{v}, \ion{Si}{vii}, and \ion{Ne}{vii}. It
is remarkable that, without fine-tuning, our model fits the
\ion{O}{v}~1371\,\AA\ line very well (Figs.\,\ref{fig2}, \ref{fig3}), indicating
that, in conjunction with the fit to \ion{O}{vi} lines, \Teff\ is now known to
an accuracy of about 3\% \citep{werner:93}.

A close inspection of the data reveals about 70 unidentified photospheric
lines. Many of them are probably \ion{Ne}{vii} multiplets, but the problem is
that their positions are rather uncertain (within several \AA). This seriously
hampers the identification of other species. In particular, the identification
of \ion{Mg}{vii} will be very difficult. Figure~\ref{fig2} shows the regions
around the three \ion{Mg}{vii} multiplets in more detail. It is not obvious that
the predicted lines have counterparts in the observation. The two strongest
predicted lines are shown in even more detail in Fig.~\ref{fig4} (top panel). No
lines are visible at the NIST wavelength positions. Like for \ion{Ne}{vii}, the
\ion{Mg}{vii} line positions are uncertain. It could well be, that some of the
unidentified features in Fig.\,\ref{fig4} are \ion{Mg}{vii} lines. For example,
shifting the computed \ion{Mg}{vii} lines by 5.5\,\AA\ (lower panel) puts them
in place of two observed lines. A systematic analysis of \ion{Mg}{vii} and
\ion{Ne}{vii} level energies will be necessary in order to obtain reliable
results.

A number of other line features are from the following potential candidates (but
often we face again problems with line position accuracy): \ion{Na}{vii},
\ion{Mg}{viii}, \ion{Si}{viii}, \ion{Ar}{viii}, \ion{Ca}{x}, and \ion{Fe}{viii}. The
reality of many features is confirmed by the presence of similar features in a
high-resolution, high-S/N \emph{HST/STIS} E140H archival spectrum of the
PG1159-type central star of NGC\,246 (\Teff\,=\,150\,000\,K, \logg\,=\,5.7
\cite{werner:06}).

\vspace{3mm}\noindent
{\bf Acknowledgements} \quad TR is supported by German Aerospace Center
(05\,OR\,0806).

\end{document}